\documentstyle[prl,multicol,aps,epsf]{revtex}
\begin{document}
\title{ Spin-$\frac{1}{2}$ Collective Excitations in BEC of 
Interacting Spin-1 
Atoms}
\author{F. Zhou}
\address{ITP, Minnaert building, Leuvenlaan 4,
3584 CE Utrecht, The 
Netherlands}
\date{\today}
\maketitle

\begin{abstract}
We construct
spin-$\frac{1}{2}$ collective excitations in BEC of interacting spin-1 
atoms. 
These excitations
exist in states with a maximal
global degeneracy.
The stability and energy of these objects are determined
by interactions with spin fluctuations and are
studied based on
a duality relation
between hyper-monopoles and magnetic monopoles in physical space.

\end{abstract}


\begin{multicols}{2}

\narrowtext

\vspace{-4cm}

Theories for Bose-Einstein condensates(BEC) of 
interacting spin-1 atoms
were recently investigated  
by a few groups\cite{Ho,Ohmi,Law,Castin,Zhou,Demler}, 
motivated by experiments on spinor BEC\cite{Stamper-Kurn}.
Of particular interest is spinor BEC with two-body scatterings
dominated by the total spin equal to 2 channel, i.e., with
an antiferromagnetic interaction.
Compared with BEC of spinless atoms,
this problem poses greater challenge to theorists and 
attracts considerable interest because of the following 
aspects. First,
the ground state of noninteracting spin-1 bosons, 
once condensed, has a spin degeneracy proportional to
the square of the number of bosons involved.
A spin dependent two-body interaction 
should be taken into 
account for the understanding of the low energy 
properties\cite{Law,Ho,Ohmi}. 
Second, a spin long range order 
could be present in the thermal dynamical limit.
Scattering between Goldstone modes, however, 
disrupts the order in the system and
is crucial to the stability
of the rotation symmetry broken BEC state. 
Third feature here is that 
the  order parameter space is $[U(1)\times S^2]/Z_2$\cite{Zhou}. This 
entanglement 
of a unit circle and a unit sphere was shown to lead to
$Z_2$ gauge fields  and the existence of other exotic many-body states  
such as a condensate of singlet pairs\cite{Demler}.

Though a qualitative understanding of the nature of many-body ground 
states in spinor BEC with antiferromagnetic 
interactions has been achieved after these 
theoretical efforts, 
the nature of collective spin excitations hasn't been explored thoroughly 
and remains to be understood.
The purpose of this paper is to illustrate the possibility of
having spin-$\frac{1}{2}$ excitations, with their quantum
number being fractions of that of an atom in BEC.
Main results of this paper are that
1) a spin-$\frac{1}{2}$ collective excitation carries a hyper-monopole
defined in an internal space,
which can be seen as a magnetic monopole with respect to a hedgehog defect 
in spinor BEC; 2)the energy of a spin-$\frac{1}{2}$
excitation is determined by interactions between hypermonopoles and spin 
fluctuations
and can be finite in the thermal dynamical
limit.

To demonstrate the existence of spin-$\frac{1}{2}$
excitations, we are going to introduce a lattice model. At each site,
a spinor BEC of $N_0$ spin-1 atoms with antiferromagnetic
interaction is formed and is described by a Hamiltonian

\begin{eqnarray}
{\cal H}_{z.m}=
\frac{\rho g_p {N}^2}{2N_0}-\mu N
+ \frac{\rho g_s {\bf L}^2}{2N_0}.
\end{eqnarray}
Here $g_s=4\pi \hbar^2 (a_2-a_0)/3M$ ($>0$) and
$g_p=4\pi \hbar^2 (2a_2+a_0)/3M$; 
$a_{0,2}$ are the scattering lengths
of two bosons in total spin $F=0,2$ channel respectively;  
$M$ is the mass of an atom.
$N_0(>>1)$ is the number of bosons at chemical potential
$\mu=\rho g_p$.
$\rho$ is the number density of atoms;
$N$ and 
${\bf L}$ are the number and total spin operators of the spinor BEC.
Eq.1 is valid as far as
the spinor BEC can be treated as a rigid rotor.

As shown in previous papers\cite{Zhou,Demler},
${\bf L}$ can be considered as an angular momentum operator
defined on a unit sphere hyperspace $S^2$ of ${\bf n}$, a nematic 
director. ${\bf L}$ and 
${\bf n}$ satisfy a commutation relation: $[{\bf L}_\alpha, {\bf 
n}_\beta]=i\hbar\epsilon_{\alpha\beta\gamma} {\bf n}_\gamma$ 
($\epsilon_{\alpha\beta\gamma}$ is an antisymmetric tensor).
Similarly, ${N}$ is an angular momentum operator
defined in a hyperspace, a unit circle ${\exp (i\chi)}$ and $[N, 
\chi]=i\hbar$. In a representation of
$|{\bf n}>$ and $|\chi>$, eigenstates of 
collective operators ${\bf n}$ and $\chi$,
the collective excitations are characterized by a wavefunction
$\Psi({\bf n},\chi)$.
The symmetry
of the N-boson wavefunction requires 
$\Psi(-{\bf n}, \chi+\pi )=\Psi({\bf n}, 
\chi)$.
If an arbitrary state is labeled with two quantum numbers of operators 
$N-N_0$ and ${\bf L}$ as $(n, l)$, a physical excitation 
has to be of a form $(n, 2l-n)$, with $n, l$ as  
integers.

Most generally,
the Hilbert space defined at each site by the Hamlitonian
in Eq.1 consists topologically distinct sectors. To classify them,
we then introduce the following global gauge transformation
$|{\bf n}>\rightarrow g_{h}({\bf n}) |{\bf n}>$,
$|{\chi}> \rightarrow g_{v}({\chi}) |{\chi}>$.
The first one defines an $S^2$ 
fiber bundle over a unit sphere
$S^2$ (the base space), i.e., $E: S^2 \times S^2$ specified by a mapping 
$g_h({\bf n}) : S^2 \rightarrow S^2$.
The second one is for a usual $U(1)$ bundle
over a unit circle specified by a mapping  
$g_v(\chi): S^1 \rightarrow S^1$.   
In both cases, gauge fields can be topological
nontrivial. Namely they are characterized by hyper-monopoles and hyper-flux
lines with charges given by winding numbers $q_{h,v}$

\begin{eqnarray}
&&q_h=\frac{i}{2\pi}\int^\pi_0 d\theta \int^{2\pi}_{0}
d\phi \epsilon^{\theta_1\theta_2} Tr \big( \partial_{\theta_1} 
g_h^{-1}({\bf n}) 
\partial_{\theta_2} g_h({\bf n}) \big), \nonumber \\
&& q_v=\frac{i}{2\pi}\int_0^{2\pi} d\chi g_v^{-1}(\chi) \partial_\chi 
g_v(\chi); 
\end{eqnarray}
here $\theta_{1,2}=\theta,\phi$ and
${\bf n}=(\sin\theta \cos\phi, \sin\theta\sin\phi, \cos\theta)$.
$Tr$ is carried over a spinor representation of $g_h({\bf n})$
(specified after Eq.3.).
Each topologically distinct
sector defined in Eq.2 is characterized by $q_h, q_v $, 
$q_{h,v}=0,\pm 1, \pm 
2$,....

${\bf L}$ and ${N}$ can be expressed as differential
operators defined in hyperspace $S^2$ and $S^1$.
Particularly, in sector $q_h, q_v $,
the gauge invariant forms of ${\bf L}$ and $N$ are

\begin{equation}
{\bf L}={\bf n} \times \big[i\frac{\partial}{\partial {\bf n}}-
{\bf A}^h({\bf n})\big], {N}=i\frac{\partial }{\partial \chi}- 
A^v({\chi})
\end{equation}
where we introduce gauge potentials ${\bf A}^h({\bf n})=i Tr g^{-1}_h({\bf 
n})\partial_{\bf n} g_h ({\bf n})$, ${A}^v=ig^{-1}_v \partial_{\chi} g_v$
in hyperspaces $S^2$ and $S^1$ respectively. 
Within each sector, the gauge potentials differ by a pure gauge 
transformation which preserves the invariant defined in Eq.2.
For sector $q_h=1$, or more conveniently
a hypermonopole of unit 
charge,
the singular gauge transformation 
$g_h({\bf n})=u({\bf n})$ is defined as a spinor in a 2d 
representation: $\big( \cos\theta/2 \exp(-i\phi/2)$, $\sin\theta/2 
\exp(i\phi/2) \big)^T$, corresponding to have two half Dirac strings 
inserted at
the south and north pole of the unit sphere.
The vector potential on the unit sphere  
${\bf A}^h=\frac{1}{2}
({\bf n}\times {\bf e}_z)({\bf n} \cdot {\bf e}_z) /
|{\bf n} \times {\bf e}_z|^2$. 
For a sector
$q_v=1$, or a hyperflux of unit charge, the gauge potentials
are $A^v=1$ and ${\bf A}^h=0$.

Following Eqs.1-3, it is clear that in the absence 
of a hypermonopole, one can choose ${\bf A}^h=0$ and redrive
the excitation spectrum obtained in previous work.
However, in sector $q_h=1$
where a hypermonopole is inserted,
the angular momentum is quantized at half integers as in the Dirac
string problem\cite{Dirac}.
Spin-$\frac{1}{2}$ excitations 
emerge as a 
result of the insertion and their wavefunctions
are
\begin{equation}
\Psi_{-\frac{1}{2}}({\bf n})=\sin\frac{\theta}{2} 
\exp[-i\frac{\phi}{2}],
\Psi_{\frac{1}{2}}({\bf n})=\cos\frac{\theta}{2} 
\exp[i\frac{\phi}{2}].
\end{equation}

These exotic excitations are surely forbidden when each site is decoupled
from each other, following the consideration of the symmetry of N-boson
wavefunctions at each site. An individual hypermonopole 
does not exist in this limit; say it differently, it is always paired 
with an anti-hypermonopole to form a bound state within each site.
However, when coupling between the BEC at different
sites is finite, the symmetry of bosonic 
wavefunction
should be imposed only globally instead of locally at each site.
Meanwhile, the total spin is a conserved quantum number and 
sectors such as $q_h=1$, which are eigenstates of the total spin,  
should be taken into account for the study of spin excitations.
Thus, it becomes possible to have a hypermonopole localized at one site 
(or a region of a few sites) which is spatially well separated from the 
antihypermonopole
at the other site.  In a region where the hypermonopole is pinned,
a collective excitation carring precisely one half spin 
appears.
The interaction between spin-$\frac{1}{2}$ excitations
is mediated by spin 
fluctuations. 
We should prove in a $3d$ lattice, 
the interaction between
these spin-$\frac{1}{2}$ excitations can be weak and
these collective excitations are well defined objects.

The estimate of the stability of a spin-$\frac{1}{2}$ excitation and the 
energy
is based on a duality between hypermonopoles and "magnetic" monopoles
in real space. 
It is convenient to introduce
a connection field ${\bf A}({\bf r})
=\epsilon^{\alpha\beta}{\bf 
e}_\alpha \nabla {\bf e}_\beta$ in real space, 
where ${\bf e}_\alpha$ and 
${\bf n}$ form
a local orthogonal basis; ${\bf 
e}_\alpha \cdot {\bf e}_\beta=\delta_{\alpha\beta}$ and ${\bf e}_\alpha 
\cdot {\bf n}=0$. The field strength is
a gauge invariant Pontryagin density ${\bf H}_k=\epsilon^{ijk} 
\partial_j A_i$ $=\epsilon^{ijk}$ $\epsilon^{\alpha\beta\gamma}{\bf 
n}_\alpha$ $\partial_i {\bf n}_\beta$$\partial_j {\bf n}_\gamma$.
A magnetic monopole carrying 
charge $Q_m$ located at ${\bf R}^m$ is defined as 
a topologically distinct configuration of ${\bf n}({\bf r})$ 
where

\begin{equation}
Q_m = \oint_{\cal S} d{\bf r} \cdot {\bf H}, \hspace{0.4cm}
\mbox{or} \hspace{0.2cm} \nabla \cdot H=Q_m\delta({\bf r}-{\bf R}^m);
\end{equation}
$Q_m=\pm 1, \pm 2,...$. 
All configurations with a monopole $Q_m$ located at
${\bf R}^m$ differ only by a 
gauge transformation in the connection field
${\bf A}({\bf r}) \rightarrow {\bf A}({\bf r}) + g_m^{-1}({\bf 
r})\nabla 
g_m({\bf 
r})$, $g_m({\bf r})$ is an analytical unitary function.

$\Phi({\bf r}^\alpha; {\bf R}^m)$
as the many-body wavefunction in the
presence of a monopole at site ${\bf R}^m$ and a spin-$\frac{1}{2}$ 
excitation or a hypermonopole at site ${\bf 
r}^\alpha$ can be written as

\begin{eqnarray}
&& \Phi \propto {\cal P}
\sum_{Q_m=1, {\bf R}^m} \Psi_0\big(\{{\bf n}({\bf r}^\gamma)\}\big)
\Psi_{\frac{1}{2}}\big({\bf n}({\bf r}^\alpha)\big)
\big |\{{\bf n}({\bf r}^\gamma)\}
\big>
\nonumber \\
&&
\big |\{{\bf n}({\bf r}^\gamma)\}
\big>=
g_h\big({\bf n}({\bf r}^\alpha)\big)
\prod_{\gamma=\beta, \alpha}
\prod_{l=1}^{N_0}u_l\big({\bf n}({\bf r}^\gamma)\big)
v_l\big({\bf n}({\bf r}^\gamma)\big)
\nonumber \\
\end{eqnarray}
The summation is carried over all configurations in
sector $Q_m=1$
with monopole located at ${\bf R}^m$.
$\Psi_0\big(\{{\bf n}({\bf r}^\gamma)\}\big)$ 
is an arbitrary single valued  
function with
$\gamma$ labeling all sites including $\alpha$;
$\beta$ labels all lattice sites expcept site $\alpha$.
We have introduced ${\cal P}$ as a permutation operator to preserve
the symmetry of the wavefunction.
In a standard three-component representation of $F=1$ state,
$uv({\bf n})$ $=\big( -\sin\theta \exp(-i\phi)/\sqrt{2}$,
$\cos\theta$, $\sin\theta \exp(i\phi)/\sqrt{2} \big)^T$.

Let us first consider
a state where
$\Psi_0\big(\{{\bf n}({\bf r}^\gamma)\}\big)$ 
$=\delta\big ({\bf n} ( {\bf r}^\gamma) -
{\bf n}^m({\bf r}^\gamma-{\bf 
R}^m)\big)$. 
Here,
${\bf n}^m({\bf r})={\bf r}/|{\bf r}|$.
Following Eq.5, this hedgehog carries a monopole  
field ${\bf H}^m({\bf r})={\bf r}/|{\bf r}|^3$ 
and satisfies Eq.5 at $Q_m=1$.
By examing Eq.6, we prove that the
Berry's phase $\Gamma_{m(h)}({\cal C})$ of moving a hedgehog
at ${\bf r}^\alpha$
(a hypermonopole at ${\bf R}^m$) around a 
hypermonopole (a hedgehog) along loop ${\cal C}$ is 

\begin{eqnarray}
\Gamma_m ({\cal C})=
-\Gamma_h ({\cal C})[\mbox{mod} 2\pi]=
\frac{\Omega_h ({\cal C})}{2}.
\end{eqnarray}
$\Omega({\cal C})$ is the solid angle spanned by loop ${\cal C}$
with respect to the hypermonopole (the hedgehog).
Eq.7 shows that hedgehogs and hypermonopoles see each other as
"magnetic" monopoles with one half
monopole strength but of opposite signs.

One should notice that the Berry's phase of moving a hypermonopole
(a hedgehog) along path ${\cal C}$ vanishes identically in the absence of
a hedgehog (a hypermonopole).
It also follows that the wavefunction of two hypermonopoles at 
${\bf 
r}^{\alpha_1}$ and ${\bf r}^{\alpha_2}$ is symmetric 
with respect to 
interchange of positions,

\begin{equation}
\Phi({\bf r}^{\alpha_1}, {\bf r}^{\alpha_2})
=\Phi({\bf r}^{\alpha_2}, {\bf r}^{\alpha_1}).
\end{equation}
The exotic excitation therefore carries spin-$\frac{1}{2}$ but obeys
bosonic statistics\cite{t'Hooft}.

In a $2d$ lattice,
a Skyrmion texture can be defined by choosing $\Psi_0\big(
\{{\bf n}({\bf 
r}^\gamma)\}\big)
=\delta\big({\bf 
{\bf n}({\bf 
r}^\gamma)-n}^s({\bf r}^\gamma-{\bf R}^s)\big)$ in Eq.6.
For a skymion located at ${\bf R}^s$ with
${\bf n}^s({\bf r})=(\sin\theta (|{\bf r}|)\cos\phi({\bf r}),
\sin\theta(|{\bf r}|)\sin\phi({\bf r}),  \cos\theta(|{\bf r}|))$
and a hypermonopole located at ${\bf r}^\alpha$,
the Berry's phase $\Gamma_h$ 
of moving a hypermonopole around a loop ${\cal C}$ of
radius $\rho$ centered at ${\bf R}_s$
is again connected with
$\Gamma_s$ of moving a Skyrmion around a 
hypermonopole along the same path, 

\begin{equation}
\Gamma_s({\cal C})
=-\Gamma_h({\cal 
C})[\mbox{mod} 2\pi]
=\pi\big(1-\cos\theta(\rho)\big). 
\end{equation}
We have assumed $\phi({\bf r})=\arg{\bf r}_y/{\bf r}_x$, and 
$\theta(|{\bf r}|)$ is an arbitary function whose asymptotic
behaviors are given as $\theta(|{\bf r}|=0)=0$ and $\theta(|{\bf r}|>> 
\xi_0)=\pi$. 
The Berry's phase $\Gamma_s$ in Eq.9 varies from $0$ at $\rho=0$ to
$2\pi$ at $\rho=+\infty$\cite{Skyrmion}.

Eqs. 7,9 are valid for any state 
$\Psi_0\big(\{{\bf n}({\bf r}^\gamma)\}\big)$
in Eq.6 therefore in a disordered limit, if $\Gamma_{m,s}$
are defined as the Berry's phases due to the presence of a hypermonopole.
In terms of ${\bf A}({\bf r})$ fields,
$\Gamma_h$ in Eq.7 indicates that a hypermonopole 
with velocity ${\bf v}_h$ experiences
a Lorentz force $\frac{1}{2}{\bf v}_h \times {\bf H}^m\big({\bf r}^\alpha
-{\bf R}^m(t)\big)$ in the connection
field of a hedgehog located at ${\bf R}^m$. 
$\Gamma_m$ in Eq.7 on the other hand shows that
a hedgehog with a velocity ${\bf v}_m$ 
experiences a Lorentz force $\frac{1}{2}{\bf v}_m \times 
{\bf H}^m\big({\bf R}^m(t)-{\bf r}^\alpha\big)$ 
at ${\bf R}^m$, with ${\bf H}^m$ being a monopole field 
of a hypermonopole at ${\bf r}^\alpha$.
So for the static hypermonopole,
an electric field  ${\bf E}({\bf 
r}^\alpha)=-\frac{1}{2}\partial_t {\bf A}\big({\bf r}^\alpha -{\bf 
R}^m(t)\big)$
is generated due to the motion
of a hedgehog;
${\bf A}$ is precisely the connection potential of a magnetic monopole 
field ${\bf H}^m=\nabla \times {\bf A}$.
We conclude a hypermonopole 
carries an electric charge 
with respect to potentials $\big({\bf A}_i, A_0\big)$
and
is minimally coupled to the connection 
fields: ${\bf j}_i \cdot {\bf A}_i + \rho A_0$.

At this point, we should emphasis that 
with respect to hypermonopoles, connection fields of $Q_m=\pm 1$ are 
{\em physically} distinguishable,
disregarding the $Z_2$ symmetry found in previous 
works\cite{Zhou,Demler}.
That is ${\bf H}_k({\bf n})=-{\bf H}_k({\bf n})$, upon an inversion
${\bf n}\rightarrow -{\bf n}$ and ${\bf H}(Q_m=+1)=-{\bf H}(Q_m=-1)$
\cite{Volovik}.     
It basically reflects
a conversion of a hypermonopole $q_h=1$ 
into an antihypermonopole $q_h=-1$ 
upon the inversion of ${\bf n}$, following Eq.6.

However,
a {\em homotopical} distinction between the connection fields of
$Q_m=\pm 1$ exists only when the influence of a linear 
defect is absent.
Otherwise, a hedgehog with $Q_m=+1$ transforms into a $Q_m=-1$ one
when moving around a 
$\pi$-disclination and half-vortex composite, or a $Z_2$ 
string\cite{Zhou}.
Moreover, a hypermonopole 
develops a Berry's phase $\Gamma_B$ in a matrix form 
or $\Gamma_B= 
\sigma_x \pi/2 $, with $\sigma_x$ an $x$-component of Pauli matrix acting 
on a 
$2d$ hypermonopole-anti-hypermonopole space. 
By moving a $q_h=1, Q_m=1$ 
compound shown in Eq.6 around a $Z_2$ string, one obtains
a $q_h=-1, Q_m=1$ compound. 
So in the presence of $Z_2$ strings, 
hypermonopoles of $q_h=\pm 1$, or the connection
field ${\bf H}_k$ of  
$Q_m=\pm 1$ hedgehogs becomes 
{\em homotopically} indistinguishable. 
We will neglect this influence of $Z_2$ strings
in the discussion of the stability of spin-$\frac{1}{2}$
excitations, for $Z_2$ strings are gapped in the limit that interests us.
As far as the $Z_2$ symmetry is broken,  
the connection field of $Q_m=\pm 1$   
should be considered to be physically and homotopically distinct.

The low energy dynamics of the Pontryagin field ${\bf H}_k$
or magnetic monopole $\{{\bf R}^m\}$ 
can be obtained by studying the effective theory of ${\bf n}$.
As shown in\cite{Zhou,Demler}, the low energy spin dynamics in the spinor
BEC is described by
a nonlinear sigma model ${\cal L}_s=1/2f (\partial_\mu {\bf n})^2$.
Derivative $\partial_\mu$, $\mu=0,1,2,3$
is defined in $d+1$ Euclidean space 
${\bf x}_\mu=(\tilde{\tau},\tilde{\bf r})$;
$\tilde{\bf r}={\bf r}/a$, and ${\tilde{\tau}}=i t c_s/ a$,
with $a$ as the lattice constant and $c_s$ is the spin-wave velocity.
In the current situation, 
$1/2f=\sqrt{E_T/E_{so}}$ and $E_T$ is an exchange integral 
between two 
adjacent sites, $E_{so}=\rho g_s/2N_0$ is the zero point rotation energy;  
and $c_s=a \sqrt{E_{so}E_T}$. 
In spin disordered limit, the
effective theory for the gauge field ${\bf F}_{\mu\nu}({\bf r})
=\epsilon^{\alpha\beta\gamma}\epsilon_{\mu\nu}
{\bf n}_\alpha \partial_\mu {\bf n}_\beta
\partial_{\nu} {\bf n}_\gamma$
can be obtained upon integration of short range spin fluctuations
and was shown to have a form 
${\cal L}_g=1/2 g {\bf F}_{\mu\nu} {\bf F}^{\mu\nu}$,
$g\sim \Delta_s$ being the spin gap in the excitation spectrum 
measured in units of $E_{so}$
\cite{Polyakov}. 
We will only be interested in a phase coherent but spin
disordered BEC, where the connection fields are induced
by spin fluctuations as discussed above.

A static spin-$\frac{1}{2}$ excitation, 
which is represented by $\Psi_{\frac{1}{2}}\big({\bf n}({\bf r}^\alpha)
\big )|0>$ and carries a hypermonopole,
interacts with the gauge field via $H_I=\frac{1}{2}\int d{\bf 
r}\delta({\bf r}-{\bf r}^\alpha) A_0({\bf r})$.
The energy of a spin-$\frac{1}{2}$ excitation therefore is
$E=-{L_0}^{-1} \ln \big< T \exp \big(- \frac{1}{2}\int^{L_0}_0 
H_I(\tilde{\tau}) 
d\tilde{\tau} 
\big) \big>$,
with $L_0$ the perimeter along temporal direction and being infinity.
In terms of induced connection fields,

\begin{eqnarray}
E=-\frac{1}{L_{0}} 
\ln \big< \exp \big(-\frac{1}{2}\int_0^{L_0} d{\bf x}_0 \int_0^{L_2}d{\bf 
x}_2 {\bf 
F}_{02} \big) 
\big>_{\{{\bf F}_{\mu\nu}\}}, 
\end{eqnarray}
and the average is taken over the partition function 
$Z(\{{\bf F}_{\mu\nu}\})$.
Only the y-component of the electric field defined by ${\bf F}_{02}$
contributes to the energy.
The calculation can be carried out parallel to that in 
\cite{Polyakov} and we summarize results here.

In $(2+1)d$,
merons are condensed; the connection field becomes massive and
has short range correlations.
The expectation value in Eq.10, which is equal to the Wilson loop 
integral of a gauge field, 
decays exponentially with an exponent proportional to $P L_0 L_2$;
$P=\exp(-\pi/g)$ is the 
quantum tunneling amplitude of merons.
The energy of a hypermonopole is linear in term of the size of
system in $2d$, or spin-$\frac{1}{2}$ excitations interact via a linear 
potential.

In $(3+1)d$ case, however, magnetic monopoles 
are gapped in weakly disorder limit ($g \ll 1$) because
$P$ vanishes
as $L_0$ goes to infinity.
The connection field is massless and has a quasi-long range correlation.
Hypermonopoles or spin- $\frac{1}{2}$ excitations interact with each other
only through massless photons and
spin-$\frac{1}{2}$ objects in 
BEC(spin disordered but phase remains coherent) are well defined free 
excitations. The 
energy of these excitations is comparable to the spin gap.
In strongly disordered limit( $g\gg 1$), 
the action is finite and
magnetic monopoles again become condensed;
situation becomes similar to that in $(2+1)d$.
What happens here is that as an electric charge, or 
spin-$\frac{1}{2}$ excitation 
is inserted into a 
magnetic monopole condensate, the electric flux emitted forms a flux tube.
In strongly disordered limit, 
the interaction between two spin-$\frac{1}{2}$ excitations 
is linearly proportional to the length of the electric flux tube
connecting them. 
Only when monopoles uncondense as in a weakly disordered limit
in a $3d$ lattice,
an individual spin-$\frac{1}{2}$ collective
excitation is liberated from its interaction
with the connection field.

The condensation of monopoles also implies a reduction of
the ground state degeneracy.
Monopoles interact with each other via a Coulomb force and form a 
Coulomb gas once condensed.
In a Sine-Gordon representation for the monopole  
gas, or a Coulomb gas, where the monopole charge $Q_m$ couples with 
the Sine-Gordon field $\chi_{sg}$ with a term $\exp(i\chi_{sg} Q_m)$,
the partition function $Z=\int D{\chi}_{sg} \exp(-S_{SG})$ and the action 
is $S_{SG}\sim -\int d\tilde{{\tau}}P 
\cos\chi_{sg}$ in a zero mode limit\cite{Polyakov}.
Clearly, at the mean field approximation, the ground state degeneracy is 
infinite when monopoles are gapped ($P=0$)  
and singlefold (up to a $2\pi$ phase shift)  when they 
condense ($P\neq 0$)\cite{Haldane}.
In addition, all BEC states discussed in this paper have twofold $Z_2$ 
degeneracy in a cylindrical geometry, with  
$Z_2$-vortices gapped\cite{Demler,Senthil}.
Apparently, spin-$\frac{1}{2}$ excitations exist only in
states with a maximal global degeneracy.

To conclude, 
we want to make the following remarks concerning 
the topological spin-$\frac{1}{2}$ excitations.
First, spin-$\frac{1}{2}$ excitation exists when spin is gapped and 
incompressible. For BEC of a finite size, the spin-$\frac{1}{2}$ 
excitation 
should be
understood as a point like object localized at a hypermonopole plus
a uniform spin density along the edge of the BEC so that total spin
of the system is an integer. This is similar to the charge 
fractionalization in fractional quantum Hall system\cite{Robert}.
Second,
through a Hopf projection from 
${\cal CP}^1$ complex fields of $z=(z_1, z_2)$ ($z^+z=1$)
living on $S^3$
to a unit vector ${\bf n}$ on $S^2$,  the 
NL$\sigma$M derived for spinor BEC in \cite{Zhou} can be
formulated in a complex field representation.
Spin-$\frac{1}{2}$ 
$z$-quanta interact with each other through compact $U(1)$ gauge 
fields
which have confining-deconfining phases in $(3+1)d$\cite{Polyakov}.
The explicit construction of spin-$\frac{1}{2}$ excitation shown here
suggests $z$-quanta as hypermonopoles.
Third,
spin-$\frac{1}{2}$ excitations are also believed to present in an 
antiferromagnetic spin lattice with integer underlying spins. 
For an $S=1$ 
antiferromagnetic spin chain, with Hamiltonian $H=\sum_{i}
[{\bf S}_i \cdot {\bf S}_{i+1} -\beta ({\bf S}_i \cdot {\bf S}_{i+1})^2]$,
a valence-bond solid with onefold degeneracy is expected to 
be degenerate with a dimerized state of twofold degeneracy
at $\beta=1$ point\cite{AKLT}.
The Bethe Ansatz solution at $\beta=1$ point reveals that
one-particle excitations in the spin chain are spin-$\frac{1}{2}$ 
doublets\cite{Takhtajan}. 
The mechanism of having spin-$\frac{1}{2}$ excitations in this case 
might be considered as a caricature of the
spinon deconfinement
in a Rohksar-Kivelson state for spin-$\frac{1}{2}$ antiferromagnetic 
systems\cite{Kivelson}.
In 2D frustrated antiferromagnets,
due to a Higgs mechanism,
free spin-$\frac{1}{2}$ spinons are argued 
to exist  
disregarding the underlying spins at lattice sites\cite{Sachdev}. 
The hypermonopoles discussed here appear to be intimately
connected with those fractionalized objects discovered in low dimension
integer-spin antiferromagnetic lattices.

Finally, 
a similar discussion for
$(1,0)$ or $(0,1)$ collective excitations can be carried out.
The existence of particle $(1,0)$, which carries 
$A_v=1$ unit hyperflux depends on whether the $Z_2$
gauge field is confining or deconfining, as suggested 
in\cite{Demler}. 
I would like to thank F.A. Bais, E. Demler, F. D. Haldane, G. 't 
Hooft, R. Moessner, M. Parikh, J. Striet for discussions,  
particularly S. Kivelson for sending me
his unpublished paper,
and T. Senthil for useful comments on the manuscript.

\end{multicols}

\end{document}